\documentclass[aip,jcp,showpacs,floatfix,preprint]{revtex4-1}
\usepackage{comment}
\usepackage{graphicx}
\usepackage{amsmath}
\usepackage{relsize}
\usepackage{subfigure}
\usepackage{multirow}
\usepackage{amssymb}
\usepackage{relsize}
\usepackage{dcolumn}   
\usepackage{color}
\begin{document}

\title{Graphene Oxide as an Optimal Candidate Material for Methane Storage}
\author{Rajiv K. Chouhan}\thanks{Contributed equally to this work.}
\affiliation{Theoretical Sciences Unit, Jawaharlal Nehru Centre for Advanced Scientific Research, Jakkur, Bangalore, India}
\author{Kanchan Ulman}\thanks{Contributed equally to this work.}
\affiliation{Theoretical Sciences Unit, Jawaharlal Nehru Centre for Advanced Scientific Research, Jakkur, Bangalore, India}
\author{Shobhana Narasimhan}
\affiliation{Theoretical Sciences Unit, Jawaharlal Nehru Centre for Advanced Scientific Research, Jakkur, Bangalore, India}
\affiliation{Sheikh Saqr Laboratory, ICMS, Jawaharlal Nehru Centre for Advanced Scientific Research, Jakkur, Bangalore, India }

\begin{abstract}
Methane, the primary constituent of natural gas, binds too weakly to nanostructured carbons to meet the targets set for on-board vehicular storage to be viable. We show, using density functional theory calculations, that replacing graphene by graphene oxide increases the adsorption energy of methane by 50\%. This enhancement is sufficient to achieve the optimal binding strength. In order to gain insight into the sources of this increased binding, that could also be used to formulate design principles for novel storage materials, we consider a sequence of model systems, that progressively take us from graphene to graphene oxide. A careful analysis of the various contributions to the weak binding between the methane molecule and the graphene oxide shows that the enhancement has important contributions from London dispersion interactions as well as Debye interactions and higher-order electrostatic interactions, aided by geometric curvature induced primarily by the presence of epoxy groups. 
\end{abstract}

\maketitle

\section{Introduction}
Though most vehicles today run on gasoline, there are several pressing reasons to switch instead at least partly 
to natural gas -- it is cheaper as well as cleaner burning, and reserves of natural gas are distributed more uniformly
across the globe. At present, vehicles running on natural gas make use of CNG (compressed natural gas) 
technology, involving the use of heavy, unwieldy and dangerous gas cylinders, and high pressures. The hope is to replace this by adsorptive
storage in a solid state material. However, achieving the target storage capacities and binding strengths that allow one to compete with gasoline-based technologies poses a formidable materials and engineering challenge.\cite{Dueren}

High storage capacities for methane (the primary constituent of natural gas)  have been achieved with metal organic frameworks;\cite{Dueren, Yaghi, Mason, Liu} however, 
at present these materials are rather expensive to be used at an industrial scale, and it is worth exploring other classes of materials in parallel. 
While nanostructured carbons (such as graphene and carbon nanotubes) are also attractive candidates
for use as adsorptive natural gas (ANG) sorbents,\cite{Lozano} the binding of methane to these
systems is too weak for on-board vehicular applications.\cite{Mason} The experimentally measured value of $E_{\rm ads}$,
the adsorption energy of methane,  is $\sim$12 kJ/mol on graphite,\cite{Vidali} which is well below the estimated optimal value of 18.8 kJ/mol.~\cite{Bhatia-Langmuir} 
Calculated values for $E_{\rm ads}$ on graphene range from 11--16 kJ/mol, depending on the method used.\cite{Yang,Thierfelder,Girifalco,Lu,Okamoto,Zhao,Wood}
Note that the target value for $E_{\rm ads}$ falls between the binding strengths typical of physisorption and chemisorption. 

In addition to $E_{\rm ads}$, other parameters that determine gas storage capacities  are surface area, void space and density.\cite{Morris}
Previous authors have explored the possibility of increasing the uptake of methane in carbon-based systems by, e.g.,
tuning pore size,\cite{Wilcox} and increasing porosity by pyrolysis.\cite{Lua} Here, we will focus however on just one parameter, viz.,
$E_{\rm ads}$.
Diverse strategies have been  attempted to tweak $E_{\rm ads}$ upward -- by e.g., 
edge-functionalizing by various chemical groups,\cite{Wood, Molina, Kandagal} 
activation by steam and carbon dioxide,\cite{Quinn} 
and the introduction of defects and curvature.\cite{Deb-JPCC} 
However, hitherto, none of these have succeeded in enhancing $E_{\rm ads}$ sufficiently. 
Here, we show, using {\it ab initio} density functional theory calculations, 
that replacing graphene by graphene oxide succeeds in increasing  $E_{\rm ads}$ sufficiently 
so as to meet the established~\cite{Bhatia-Langmuir} target value. 
We also show that this enhancement has important contributions from three different effects, all of which act in concert here. 
We note also that graphene oxide has the additional advantage that it is more easily synthesized by a variety of chemical routes,
compared to graphene.\cite{Murali}

There has been earlier theoretical work on hydrogen binding on graphene oxide (GO);~\cite{Wang-2009, Lee-JNanoSci2013} 
however, these studies used the GO as a substrate for metal adatoms such as Ti and Ni, which act as the binding centers for gas molecules; 
we note that adding such metal atoms raises the weight of the system.   
Experimentally synthesized graphene oxide based frameworks~\cite{Yildirim-JMC} show high isoteric heat of adsorption and enhanced uptake of H$_2$, 
due to the large surface areas of frameworks achieved by boronic acid based pillaring between GO layers.   
 
\section{Calculation Details}

Our \textit{ab initio} density functional theory (DFT) calculations have been performed using the Quantum ESPRESSO package,\cite{QE-JPCM2009} with plane wave cut-offs of 40 Ry and 400 Ry for the wavefunction and charge density, respectively, and ultrasoft pseudopotentials.\cite{vanderbilt}  
Exchange correlation interactions were treated within a generalized gradient approximation.\cite{perdew1996} 
Importantly, in order to have an accurate  treatment of weak binding at a relatively modest computational cost, we have incorporated van der Waals interactions using  the ``DFT-D2" method.\cite{Grimme2007} The Brillouin zone was sampled with grids commensurate with a ($12\times12\times1$) mesh for the primitive unit cell of graphene. Convergence was aided by using cold smearing \cite{marzari1999} with a width of 0.001 Ry. Artificially periodic images normal to the plane of the graphene sheet were separated by 20 \AA. All atomic coordinates were allowed to relax, using Hellmann-Feynman forces, with a  convergence threshold of $10^{-3}$ Ry/bohr.  Several initial orientations of the methane molecule relative to the substrate were considered, in order to span the space of possible geometries.

The adsorption energy of methane is given by:
\begin{equation}
E_{\rm ads} = -(E_{\rm sub+CH_4} - E_{\rm sub} - E_{\rm CH_4}),
\end{equation}
\noindent
where $E_{\rm sub+CH_4}$, $E_{\rm sub}$ and $E_{\rm CH_4}$ are
the calculated  total energies of the substrate with adsorbed methane, the substrate alone, and an isolated methane molecule in the gas phase, respectively.

\section{Results}
First, we consider methane adsorption on bare graphene; we label this configuration C(I). We obtain $E_{\rm ads}$ = 14.37 kJ/mol; this falls within the range of previous theoretical values.\cite{Yang,Thierfelder,Girifalco,Lu,Okamoto,Zhao,Wood}  

Next, we wish to consider the adsorption of methane on graphene oxide. The structure of graphene oxide continues to be a matter of debate, and several models exist for it. The carbon atoms are arranged in the honeycomb lattice characteristic of graphene. However, hydroxyl (-OH) and epoxy {(-O-)} groups are not attached in a periodic manner to
these carbon atoms, so that the overall structure lacks crystalline symmetry.~\cite{Lerf1,Lerf2}

\begin{figure}[t]
\begin{center}
\subfigure[\ $(2\sqrt 3 \times \sqrt 3)$ PGO top-view]{\includegraphics[ clip=true, width=3.5cm]{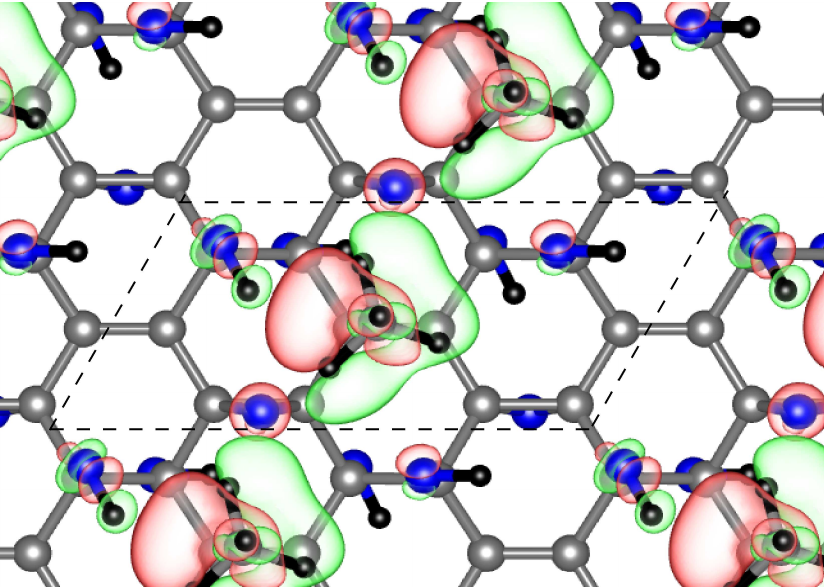}}
\quad
\subfigure[\ $(2\sqrt 3 \times \sqrt 3)$ PGO side-view]{\includegraphics[ clip=true, width=3cm]{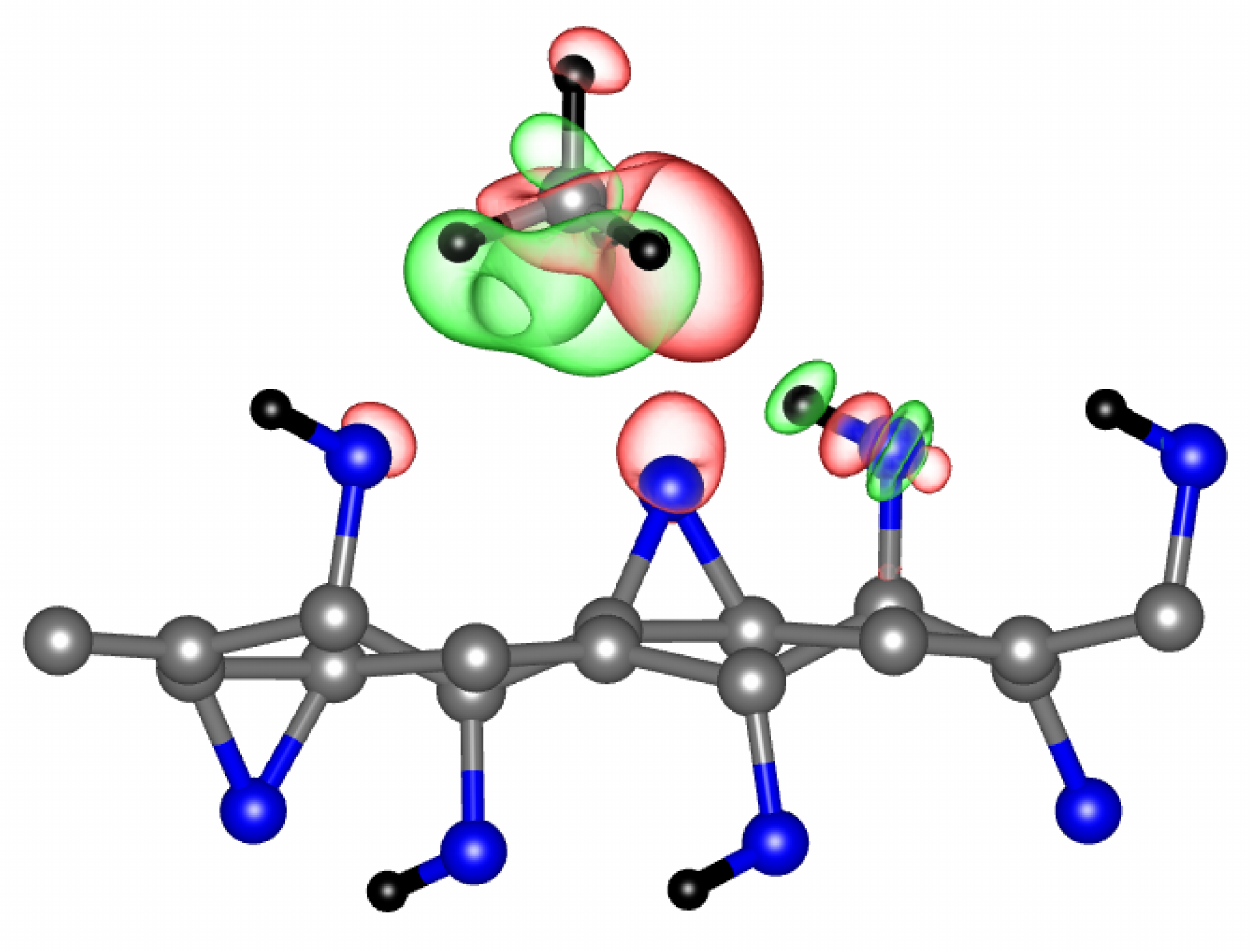}}
\quad
\subfigure[\ $(2\sqrt 3 \times 2\sqrt 3)$ PGO top-view]{\includegraphics[ clip=true, width=3.5cm]{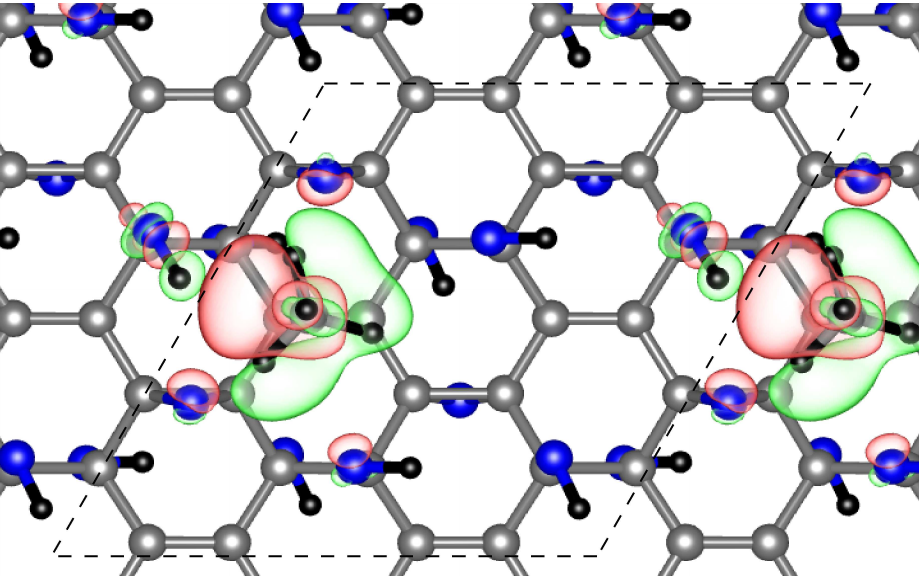}}
\quad
\subfigure[\ $(2\sqrt 3 \times 2\sqrt 3)$ PGO side-view]{\includegraphics[ clip=true, width=3cm]{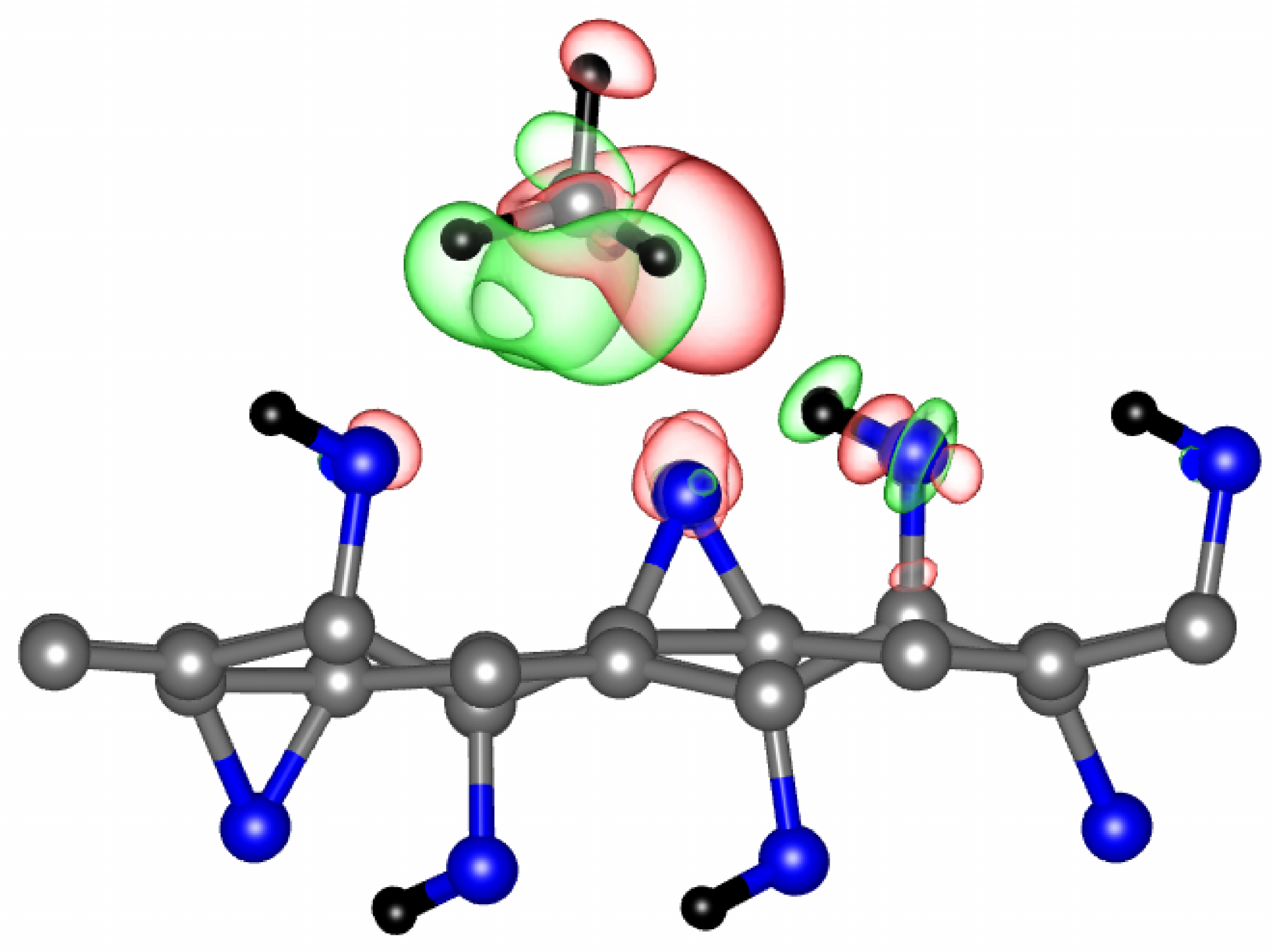}}
\quad
\subfigure[\ Z-INNER top-view]{\includegraphics[ clip=true, width=3.5cm ]{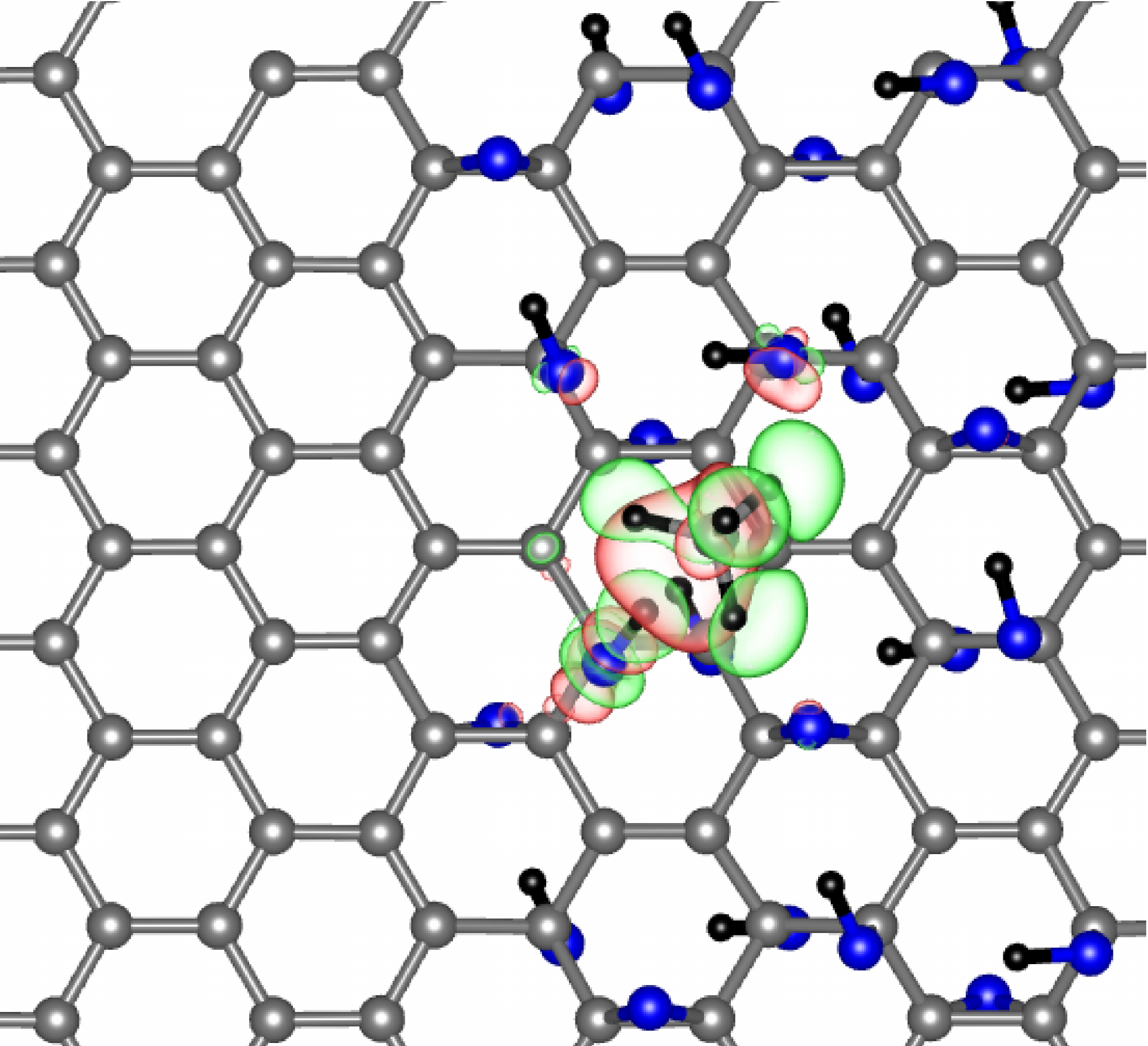}}
\quad
\subfigure[\ Z-INNER side-view]{\includegraphics[ clip=true, width=4cm ]{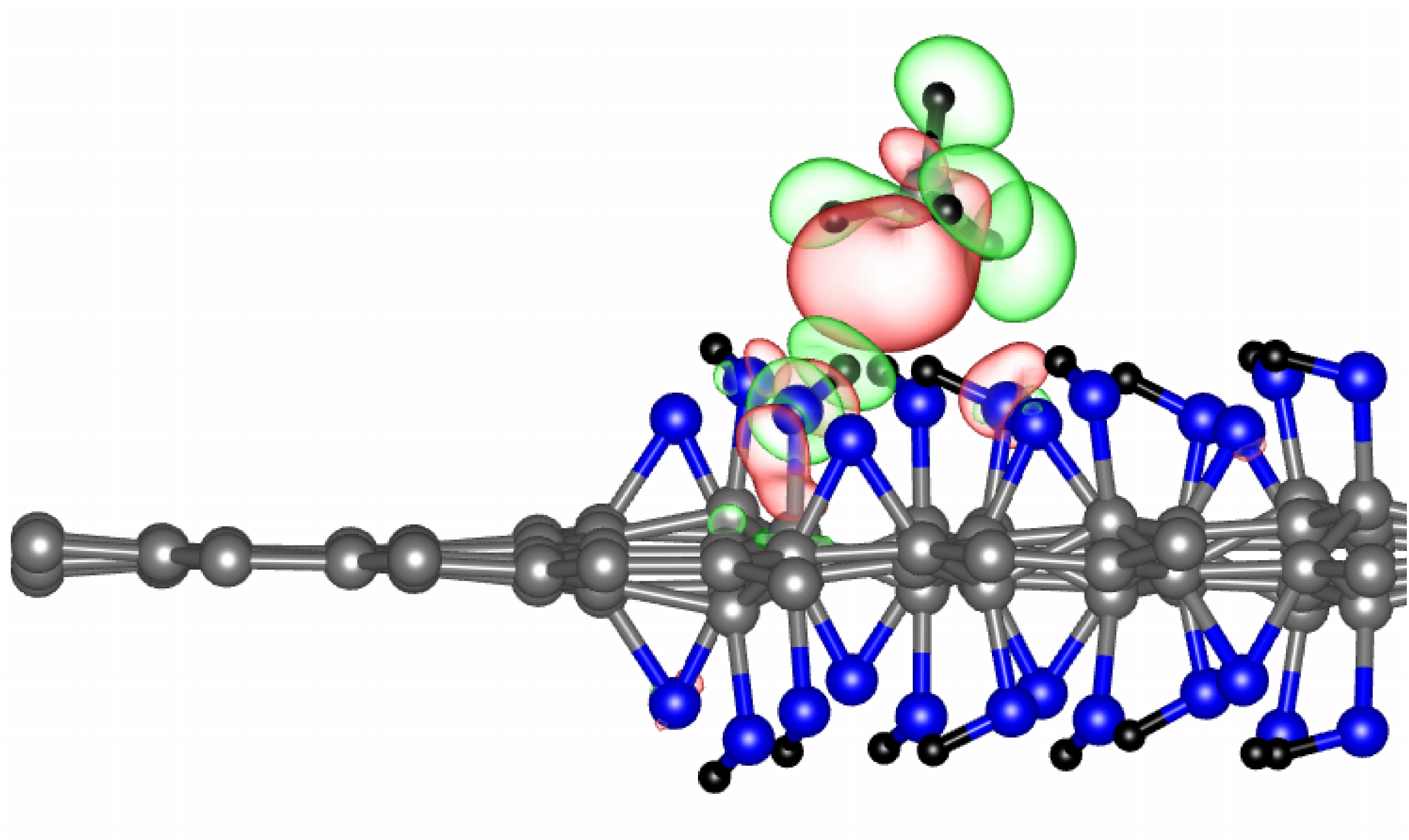}}
\quad
\subfigure[\ Z-OUTER top-view]{\includegraphics[ clip=true, width=3.5cm ]{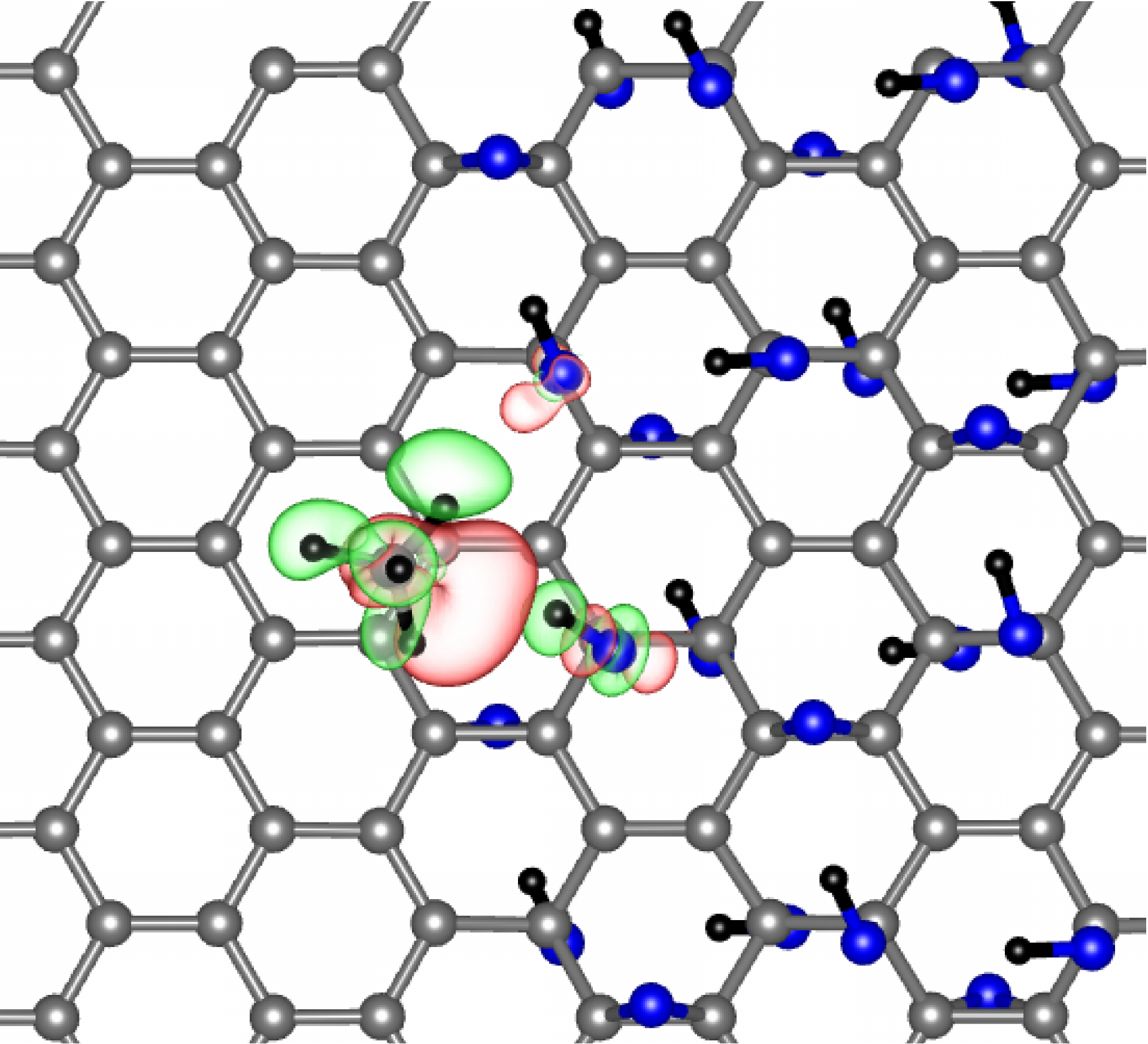}}
\quad
\subfigure[\ Z-OUTER side-view]{\includegraphics[ clip=true, width=4cm]{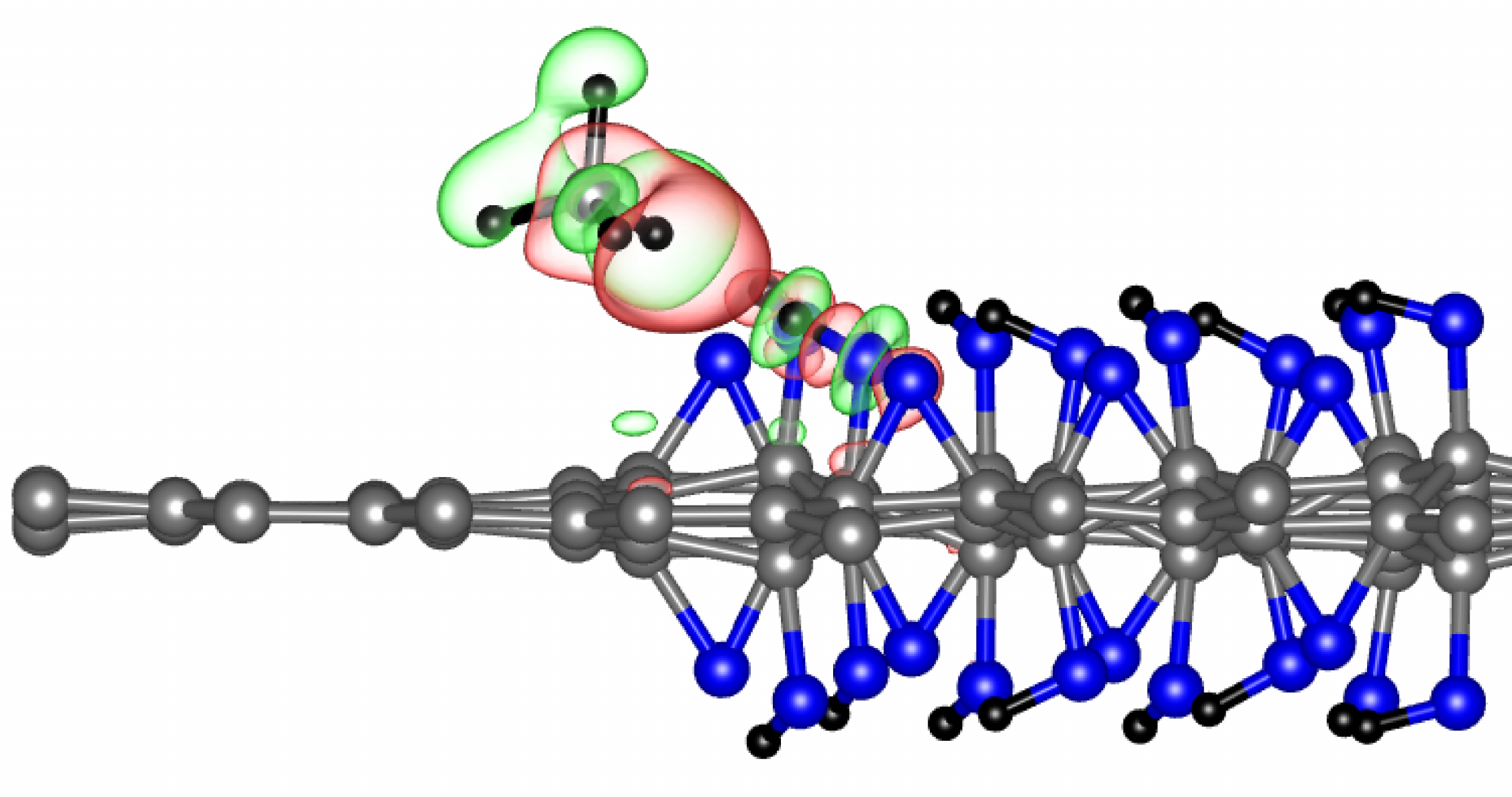}}
\quad
\subfigure[\ A-OUTER Top view]{\includegraphics[ clip=true, width=3.cm ]{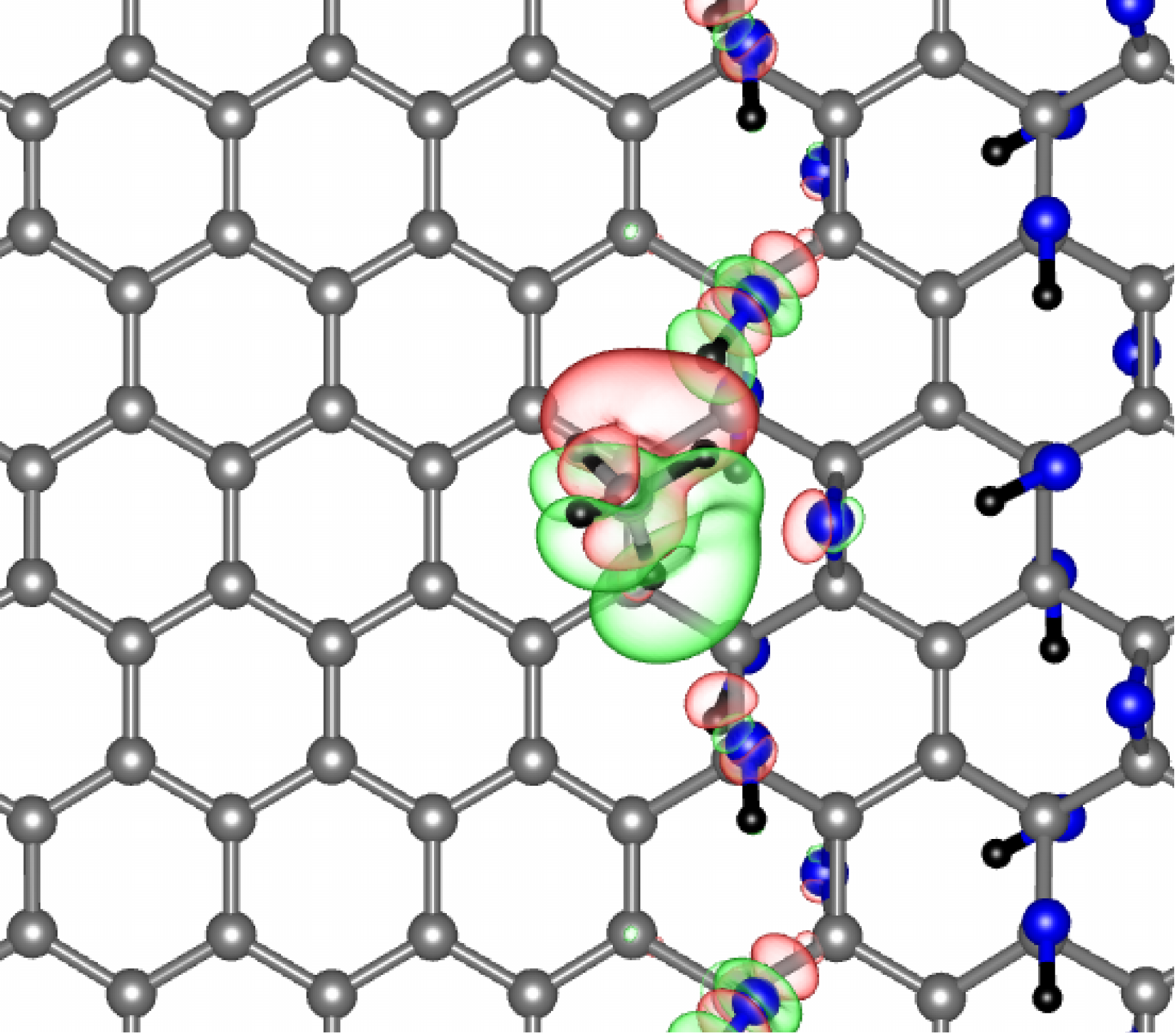}}
\quad
\subfigure[\ A-OUTER side-view]{\includegraphics[ clip=true, width=4cm]{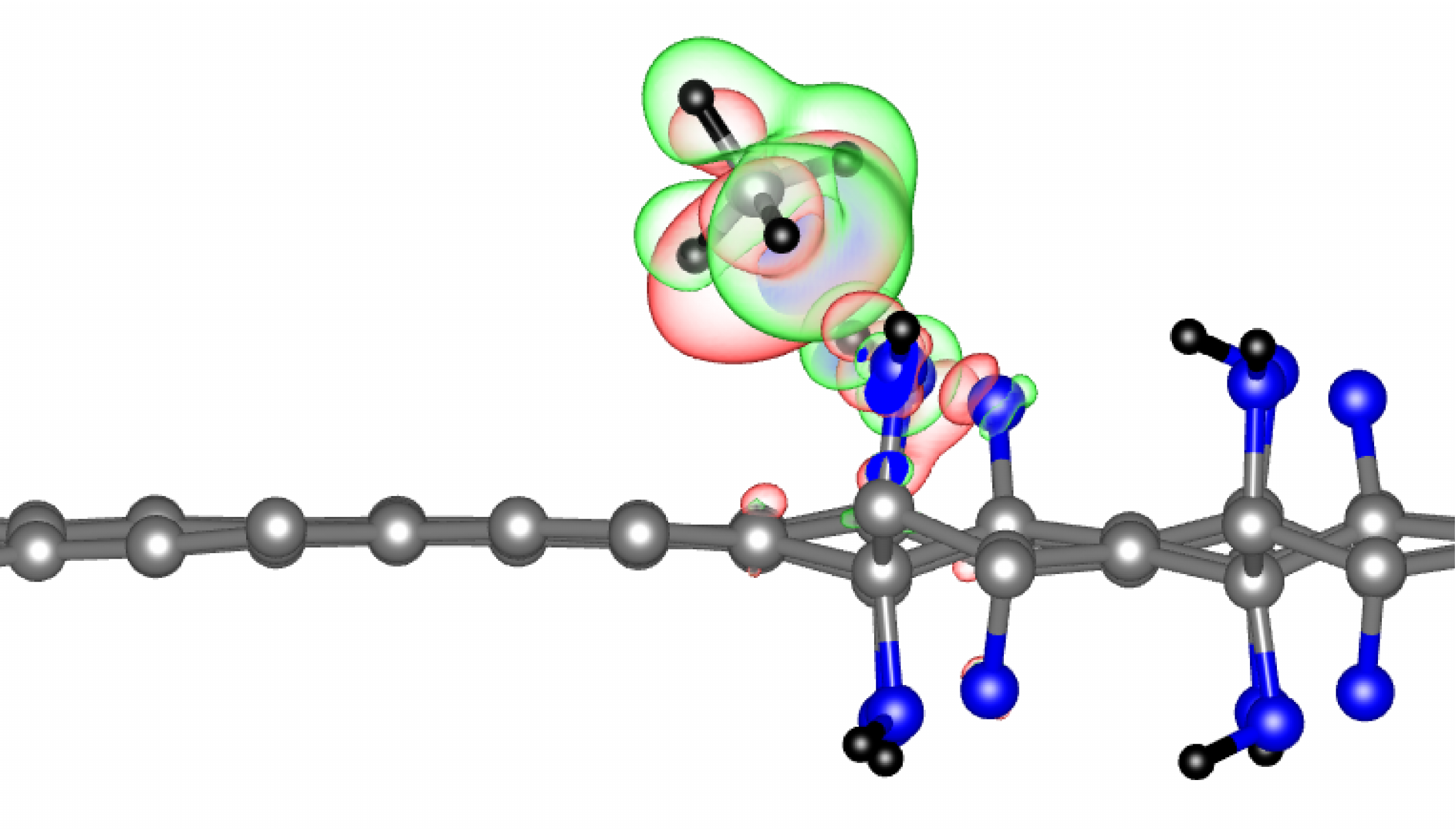}}
\caption{ Optimal relaxed geometries and  charge redistribution $\Delta\rho$ for methane adsorbed on
(a)--(d) PGO and (e)--(j) at an interface between bare graphene and PGO. Isosurfaces of $\Delta\rho= \pm 2.8 \times 10^{-4}$ $e/$bohr are plotted,
Red and green lobes denote the gain and depletion of  electronic charge, respectively.
Color scheme for atomic spheres: H (black), C (gray), O (blue). }
\label{fig1}
\end{center}
\end{figure}

We first wish to examine the energetics of methane adsorption on areas of GO that are densely covered with functional groups. 
To do this, we make use of the periodic graphene oxide (PGO) approximant of Wang {\it et al.}, which features a repeating motif of a triplet
of functional groups (two -OH on one side of the graphene sheet, and one -O- on the other side) attached to a sixfold carbon ring; this model
was motivated by the results of nuclear magnetic resonance experiments,\cite{Cai}  and guided by the energetics of various configurations as obtained from DFT.
The optimal geometry obtained
when a single methane molecule is adsorbed within the $(2\sqrt 3 \times \sqrt 3)$ primitive unit cell of PGO is shown in Figs.~\ref{fig1} (a) and (b); it leads to $E_{\rm ads}$ = 17.04 kJ/mol. 
In the more dilute case, where a single methane molecule is adsorbed instead in a $(2\sqrt 3 \times 2\sqrt 3)$ cell [see Figs.~\ref{fig1} (c) and (d)], $E_{\rm ads}$ decreases to 15.71 kJ/mol, due to a lessening of the
attractive interaction between methane molecules in neighboring cells. 

While these values of $E_{\rm ads}$ do not quite meet the targets~\cite{Bhatia-Langmuir} considered desirable for on-board storage, we must remember that the PGO approximant is not a realistic model of GO, since it describes a substrate that is densely covered with functional groups, whereas graphene oxide is believed, in reality, to contain in addition, considerable amounts of patches
of bare graphene, with many interfaces present between covered and bare areas.\cite{Nguyen-PRB, Erickson-AdvMater, Pacile-Carbon} 
Accordingly, we next consider adsorption of methane at an interface between PGO and bare graphene. We model this interface as either a zigzag or armchair-like line. -OH groups near the interface now have considerable orientational freedom, and are in fact found to re-orient themselves, compared to their alignments in the PGO structure. 

For a methane molecule adsorbed in a $(4 \sqrt 3 \times 6)$ cell containing a zigzag interface, 
we find two stable adsorption geometries that are very close in energy, we refer to these as ``Z-INNER"
[see Figs.~\ref{fig1} (e) and (f)] and ``Z-OUTER"  [see Figs.~\ref{fig1} (g) and (h)]; in these two configurations the methane molecule is positioned
near the interface, but over covered and bare regions of graphene, respectively. Most interestingly, adsorption near the interface
increases the binding, pushing $E_{\rm ads}$ up to 21.70 kJ/mol and 21.59 kJ/mol for the two geometries, respectively. 
Similarly, for adsorption near an armchair interface in a $(2 \sqrt 3 \times 4 \sqrt 3)$ cell 
[see Figs.~\ref{fig1} (i) and (j), labeled ``A-OUTER"], 
the value of $E_{\rm ads}$ is again increased, this time to 20.87 kJ/mol.
These three values of $E_{\rm ads}$ all meet the target set for on-board storage to be viable,\cite{Bhatia-Langmuir} suggesting that graphene oxide may be a
good candidate material for adsorptive storage of natural gas. Note also that by going from graphene to graphene oxide, we have raised the binding
strength by one and a half times.

\begin{figure*}[t]
\begin{center}
\subfigure[\ C(I) side-view]{\includegraphics[ clip=true, width=3cm]{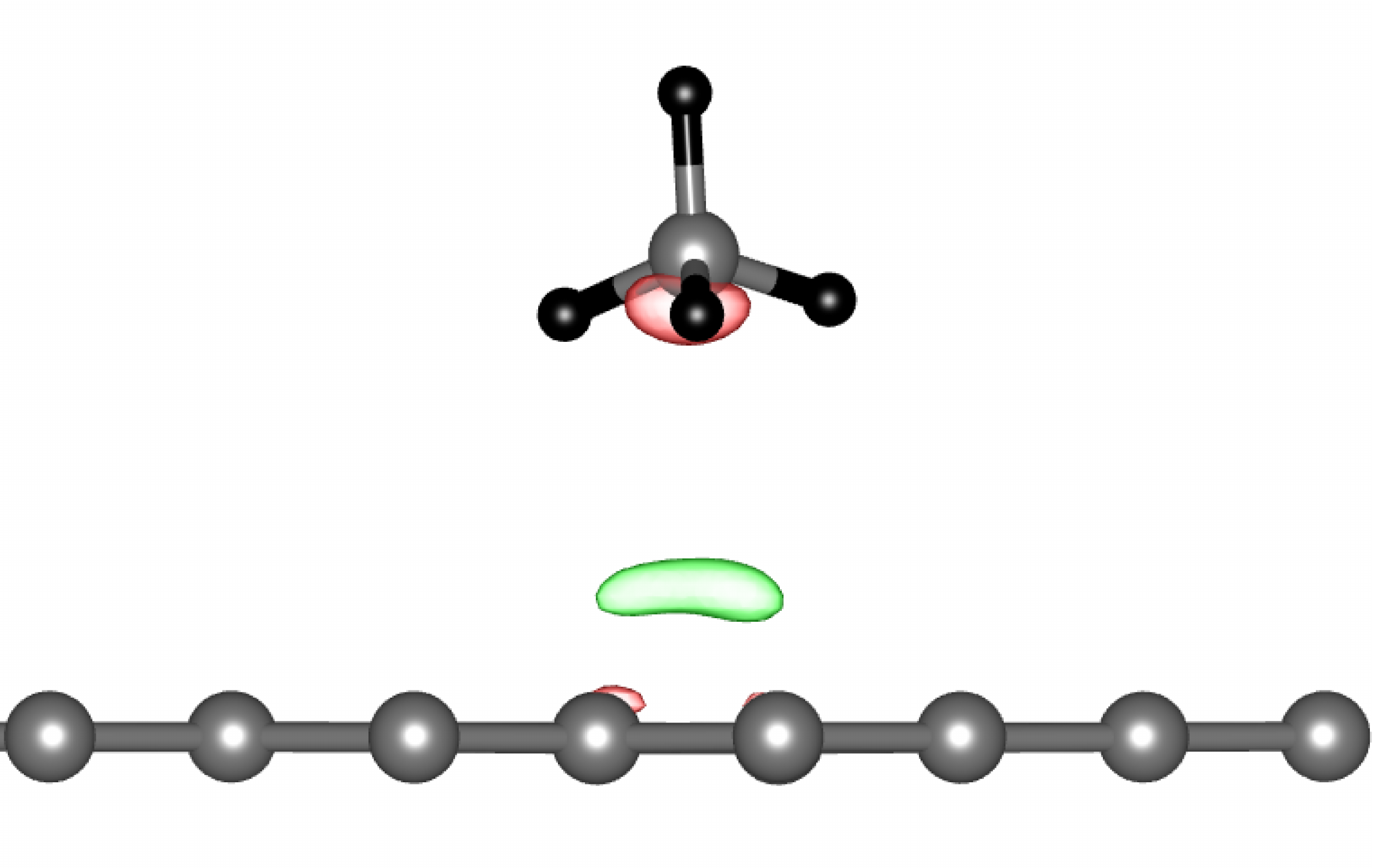}}
\quad
\subfigure[\ C(II) side-view]{\includegraphics[ clip=true, width=3cm]{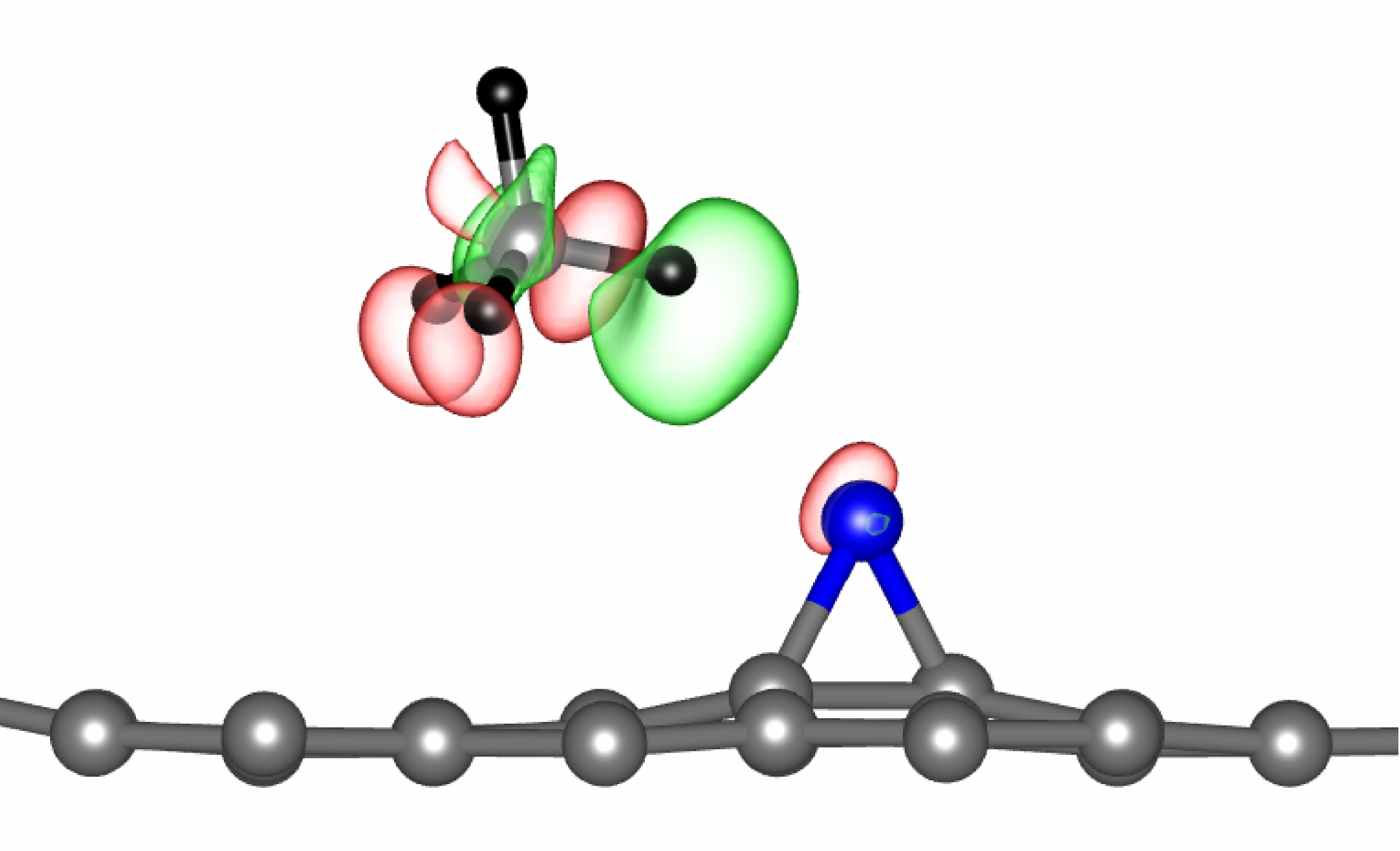}}
\quad
\subfigure[\ C(III) side-view]{\includegraphics[ clip=true, width=3cm]{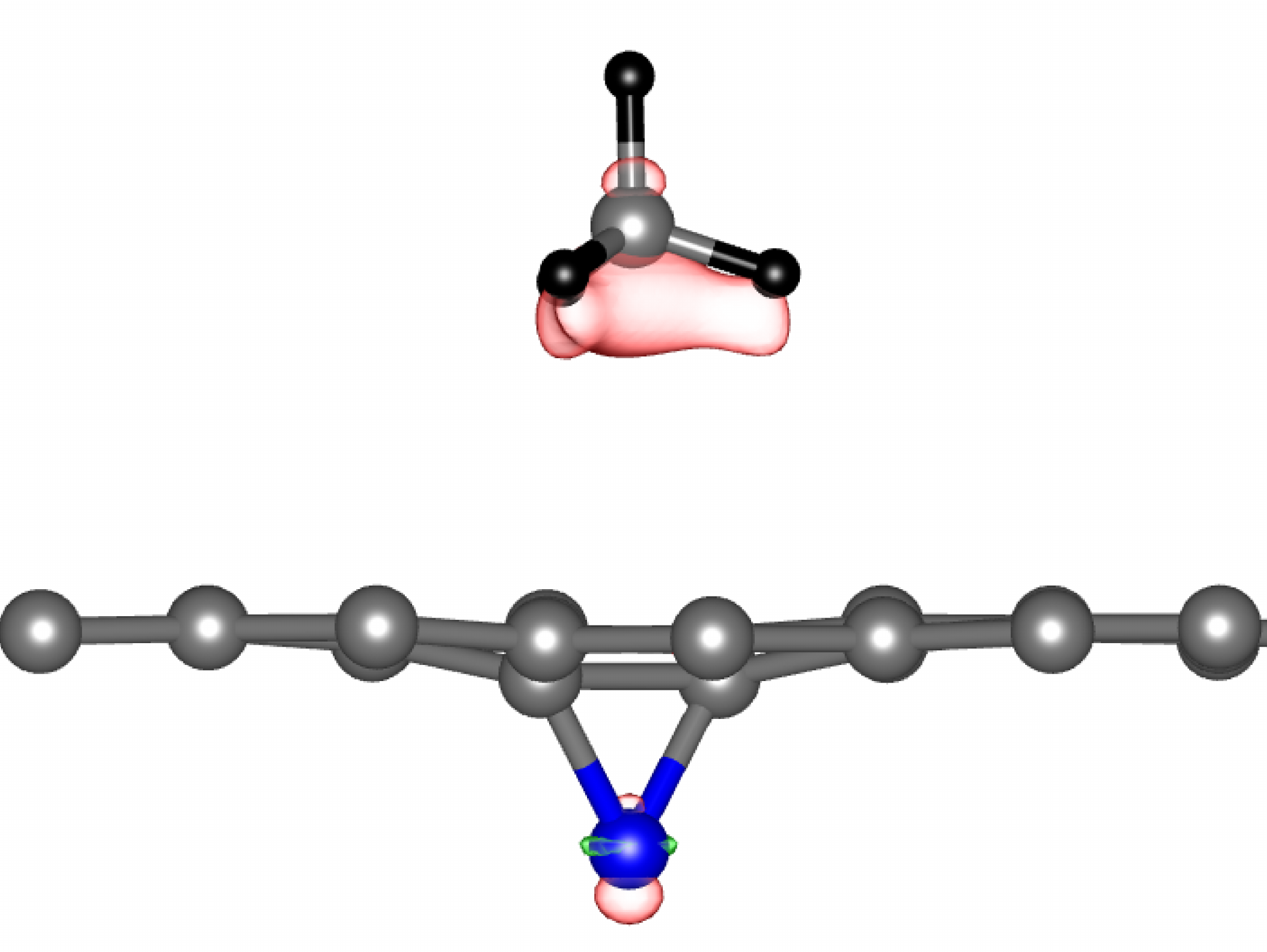}}
\quad
\subfigure[\ C(IIIa) side-view]{\includegraphics[ clip=true, width=3cm]{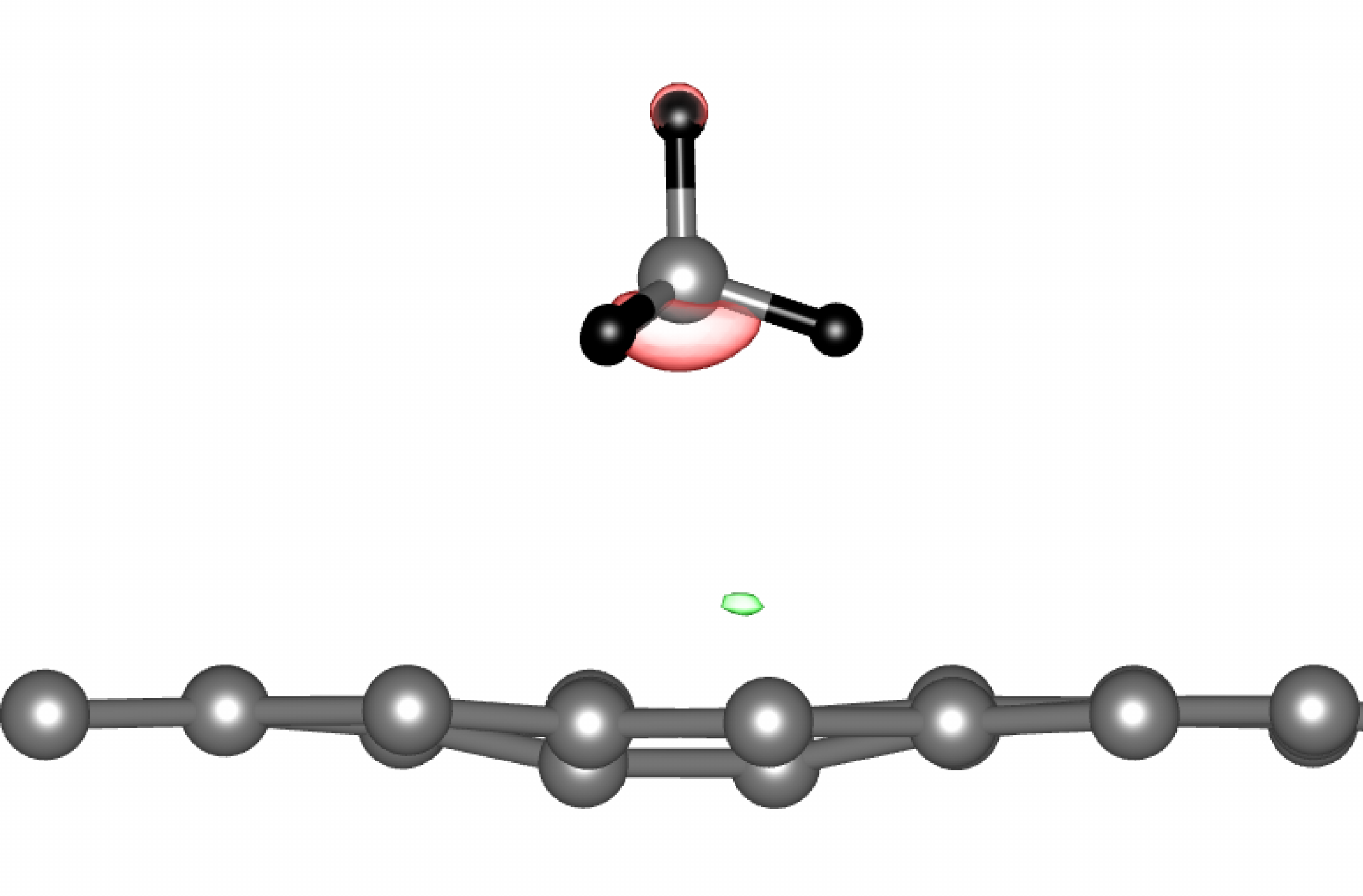}}
\quad

\subfigure[\ C(IV) side-view]{\includegraphics[ clip=true, width=2.5cm]{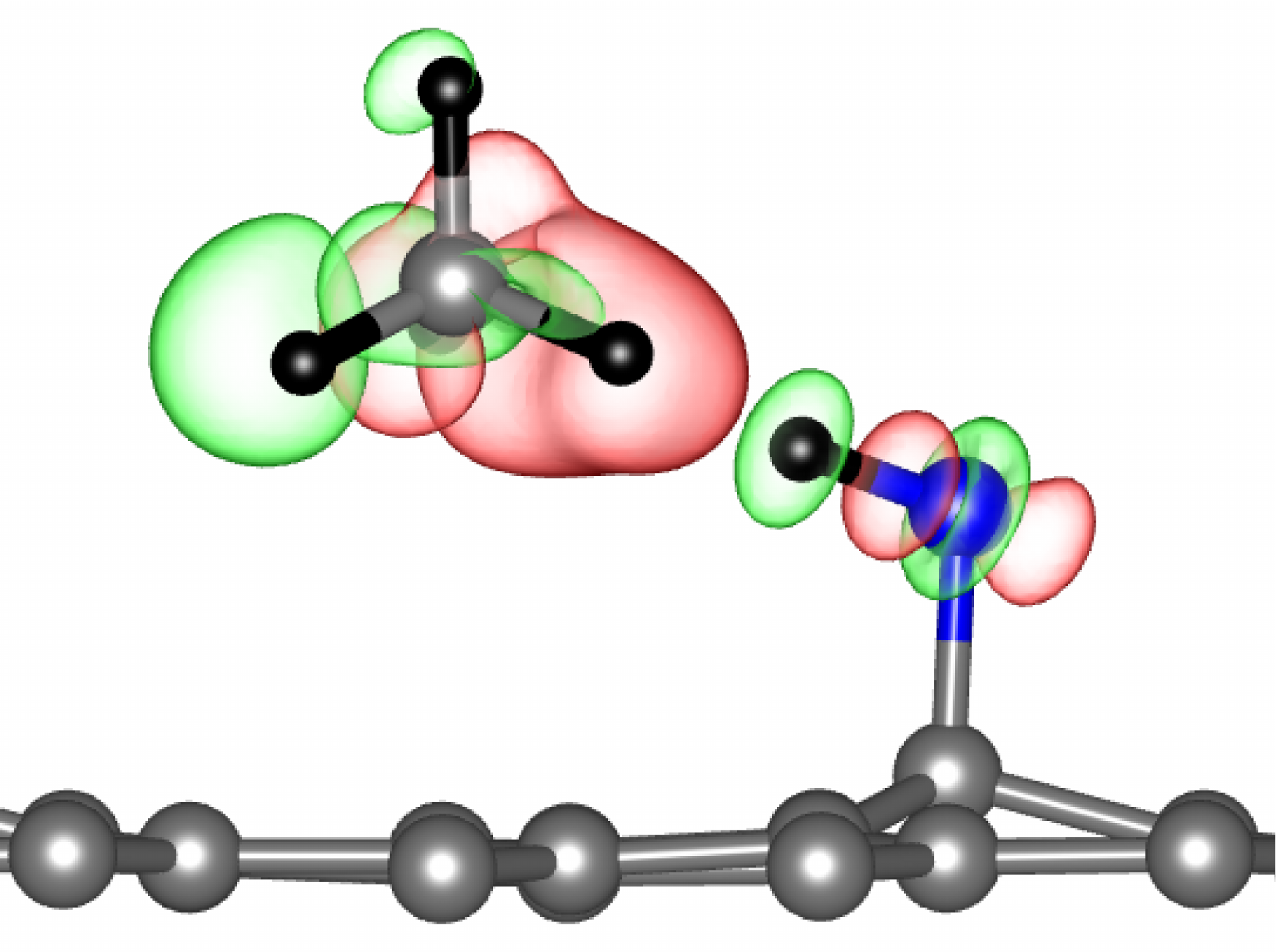}}
\quad
\subfigure[\ C(V) side-view]{\includegraphics[ clip=true, width=3cm]{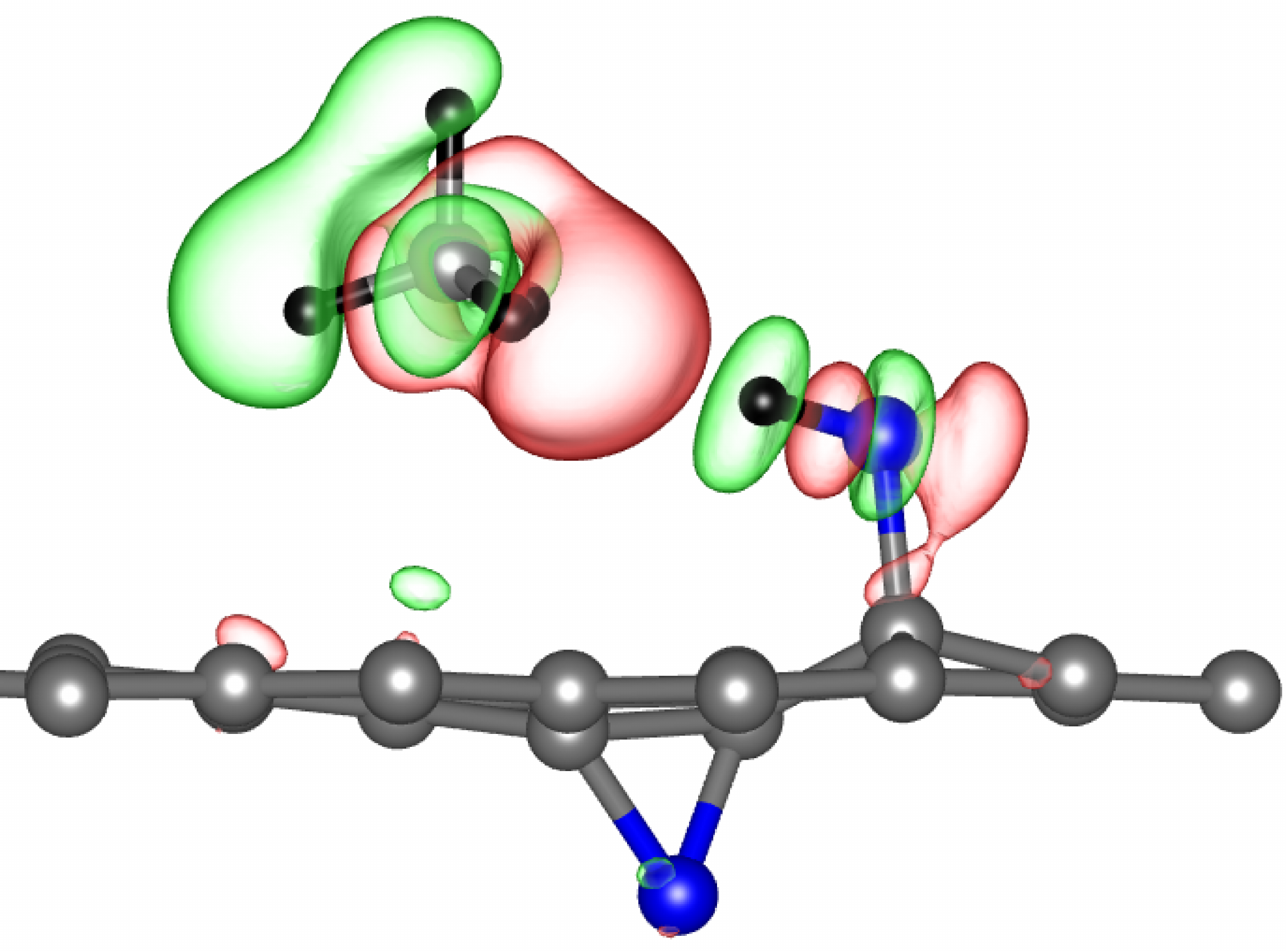}}
\quad
\subfigure[\ C(Va) side-view]{\includegraphics[ clip=true, width=3cm]{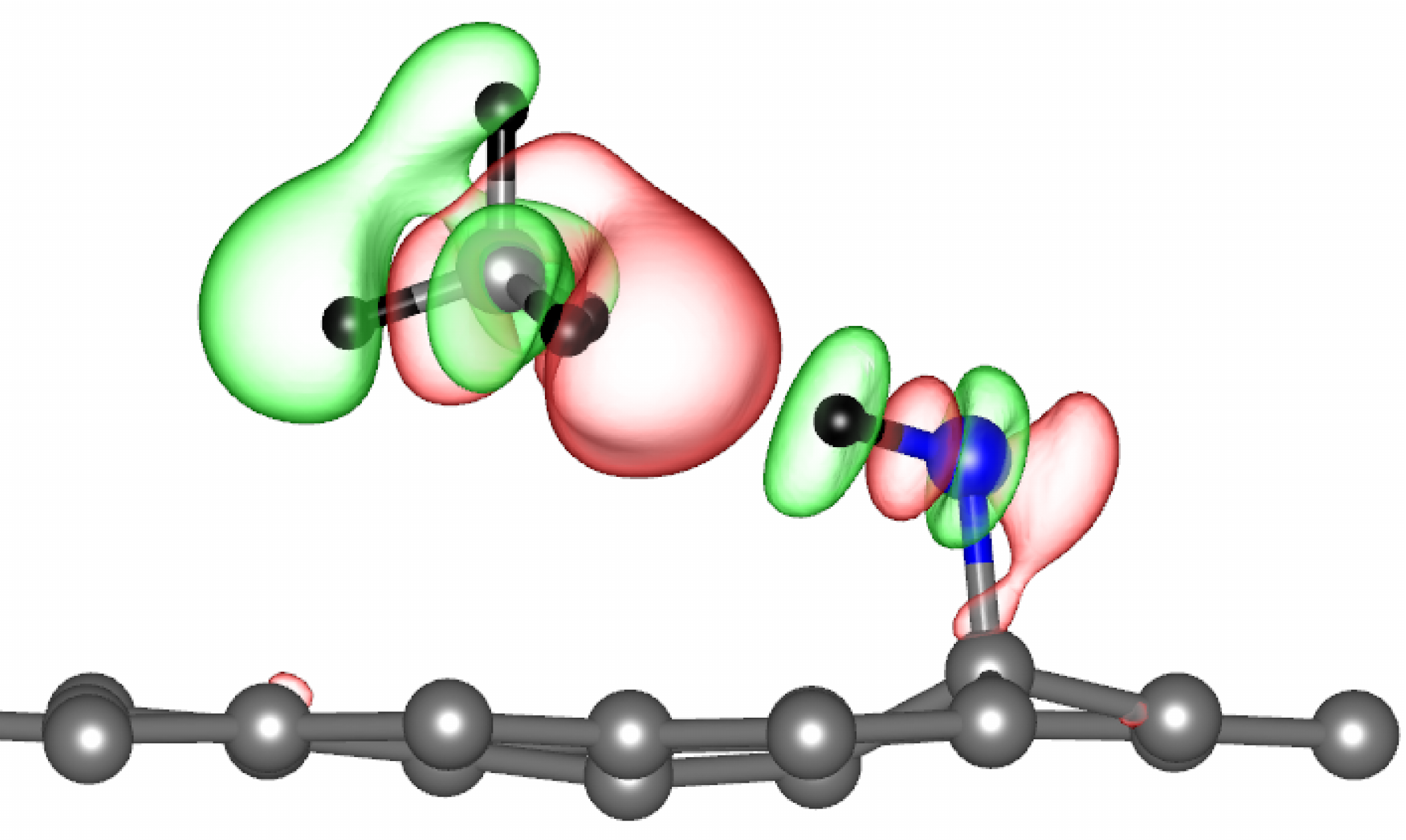}}
\quad
\subfigure[\ C(VI) side-view]{\includegraphics[ clip=true, width=3cm]{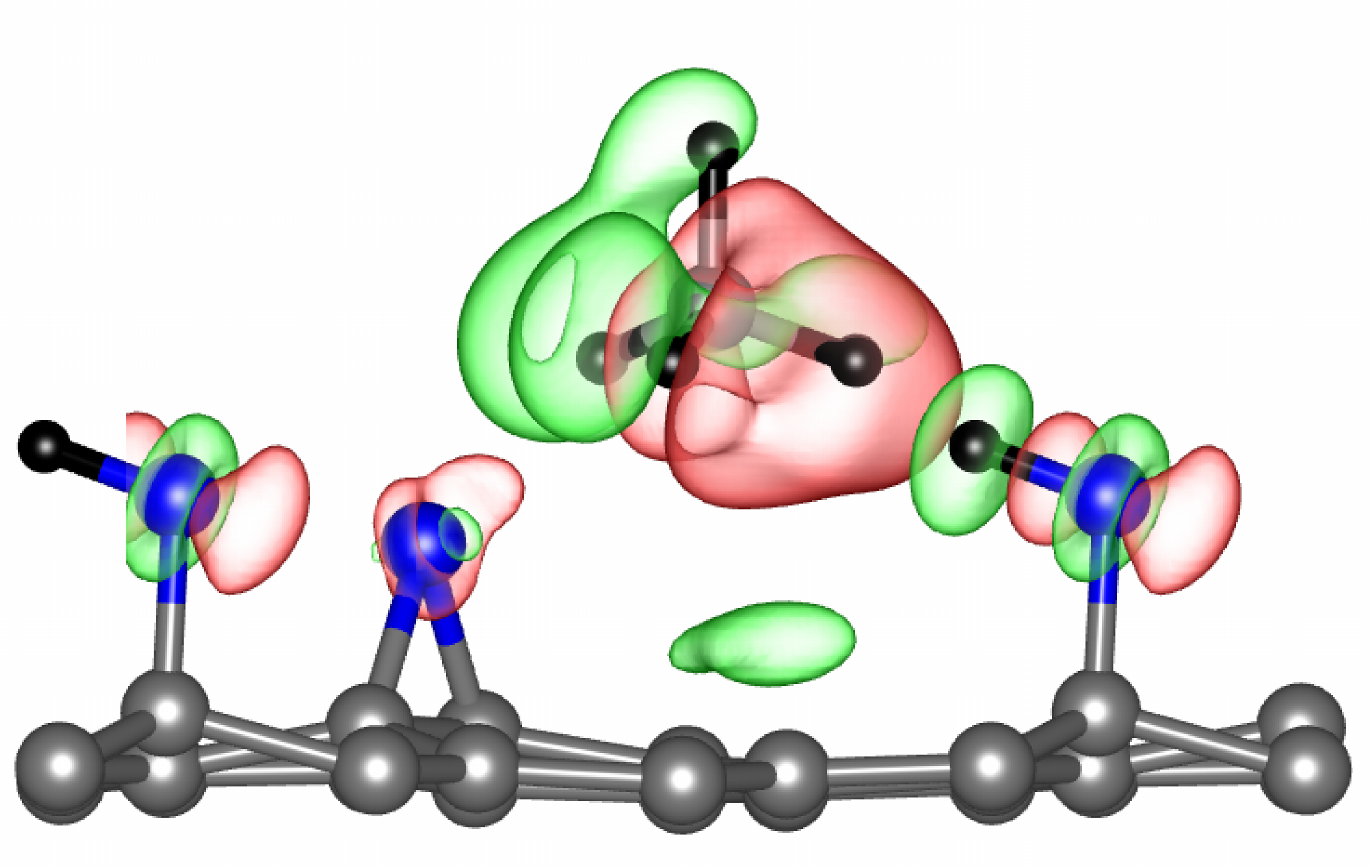}}
\caption{ Optimal relaxed geometries and  charge redistribution $\Delta\rho$ for methane adsorbed on
bare graphene [configuration C(I)] and a sequence of model systems [configurations C(II)--C(VI)]. Isosurfaces of $\Delta\rho= \pm 2.8 \times 10^{-4}$ $e/$bohr are plotted,
Red and green lobes denote the gain and depletion of  electronic charge, respectively.
Color scheme for atomic spheres: H (black), C (gray), O (blue). }
\label{fig2}
\end{center}
\end{figure*}

We now proceed to examine the origins of this enhanced binding, and try to break it down into various energetic and structural components. The main energetic contributions relevant to weak binding such as that being considered here, are London
dispersion interactions (LDI), and various electrostatic interactions (some of which are subsumed into van der Waals interactions, such as the Debye interaction between permanent and induced electric dipoles). The part of $E_{\rm ads}$ arising from LDI is given by
$E_{L} = -\big(E^{\rm disp}_\mathrm{sub+CH_4} - E^{\rm disp}_\mathrm{ CH_4} - E^{\rm disp}_\mathrm{\rm sub}\big)$,
where the three terms on the right-hand-side are the contributions from LDI to the total energies of the corresponding systems. We find that the values of $E_L$ for Z-INNER, Z-OUTER and A-OUTER are 19.08, 21.58 and 19.05
kJ/mol, respectively. This shows that the majority of the binding comes from LDI, and thus any theoretical treatment that does not account properly for these may be expected to be inaccurate. However, there are also some repulsive contributions to $E_{\rm ads}$, and the LDI is
not the only attractive interaction that offsets these. In fact, we will see that attractive electrostatic interactions between induced charges
on the methane molecule and the dipolar {-OH} and -O-  groups provide important contributions toward stabilizing the
adsorbed methane, and toward pushing $E_{\rm ads}$ into the desired window. As an illustration that interactions other than LDI are important, we note that on going from
CH$_4$ adsorbed on PGO in the $(2\sqrt 3 \times 2\sqrt 3)$ cell, to the Z-INNER interface configuration, $E_{\rm ads}$ goes up
by 5.99 kJ/mol, whereas $E_L$ is increased only by 1.37 kJ/mol.

In Fig.~\ref{fig1}, in addition to depicting the relaxed structures of the methane+substrate systems, we have also plotted isosurfaces of the charge redistribution  $\Delta \rho = \rho_{\rm sub+CH_{4}} -\rho_{\rm CH_{4}}- \rho_{\rm sub}  $, where $\rho$ is the electronic charge density, and all three terms on the right-hand-side are evaluated at the relaxed geometry of the combined system. Red and green lobes correspond to an accumulation and depletion of electronic charge, respectively. In Fig.~\ref{fig1}, we can see very clearly that though methane in the gas phase possesses neither a dipole nor quadrupole moment, when adsorbed on PGO, it displays
a considerable dipole moment. This dipole moment is larger when it is adsorbed at an interface than on PGO; a visual indication of this can be obtained by observing the larger red and green lobes in Figs.~\ref{fig1} (e)--(j) than in Figs.~\ref{fig1} (a)--(d).

In order to break down the interaction of the methane molecule with graphene oxide into various components, we now consider a sequence of
simple model systems, designed so as to progressively take us from bare graphene to graphene oxide, while allowing us to separate out the various effects at play.

We have already mentioned configuration
C(I), where methane binds to a clean
graphene sheet. In this case, the binding is almost entirely due to LDI; electrostatic interactions are 
negligible, as evidenced by the very small red and green lobes in Fig.~\ref{fig2}(a). 

Next, we consider graphene functionalized with a single epoxy group. It is interesting to note that it is far more favorable for
CH$_4$ to bind on the opposite side of the epoxy [configuration C(III), see Fig.~\ref{fig2}(c), $E_{\rm ads}$ = 17.11 kJ/mol] than on the same side [C(II), see Fig.~\ref{fig2}(b), $E_{\rm ads}$ = 12.69 kJ/mol]. This
preference has a structural origin: the attachment of the epoxy group results in a considerable local buckling of the graphene sheet, with the
two C atoms that are bonded to the O atom being displaced vertically by $\sim$ 0.24 \AA\ relative to the rest of the atoms. It is known
that methane molecules prefer to bind to a concave curvature of a graphene sheet;\cite{Deb-JPCC} this is the phenomenon being observed here. 
If we remove the O atom, but freeze the graphene sheet in the buckled configuration it assumes in the presence of the epoxy group [C(IIIa), see Fig.~\ref{fig2}(d)], 
we obtain $E_{\rm ads}$ = 15.76 kJ/mol, i.e., we recover most of the enhanced binding. As expected, C(III) and C(IIIa) have almost identical values for $E_L$. 
Note also that the isosurfaces in Figs.~\ref{fig2} (c) and (d) do not display any signature of significant electrostatic contributions to binding.

Functionalizing graphene with a single hydroxyl group leads to a quite different scenario. Methane prefers to bind on the same side as
the -OH [configuration C(IV), Figs.~\ref{fig2}(e)], and there are noticeable induced moments on the molecule, as well as some redistribution of charge on the 
substrate. The value of $E_{\rm ads}$ goes up now to 20.56 kJ/mol. Interestingly the induced dipole on methane is oriented
oppositely, with respect to the functional group, for epoxy and hydroxyl [compare Figs.~\ref{fig2}(b) and \ref{fig2}(e)].

Finally, we consider what happens when both an epoxy and a hydroxyl are present together, as in graphene oxide. In Fig.~\ref{fig2}(f) we show configuration C(V), where the
-O- and -OH are adsorbed on opposite sides of the graphene sheet in a very low energy geometry.\cite{Wang-2009}
The value obtained for $E_{\rm ads}$ is now increased further, to 21.78 kJ/mol. 
In the hypothetical configuration C(Va), we retain the -OH group of C(V), but remove the -O-; 
however, we freeze the system geometry at that of C(V) [see Fig.~\ref{fig2}(g)]. 
$E_{\rm ads}$ is almost unchanged, being equal to 21.66 kJ/mol, 
again confirming that the main role played by the epoxy group in enhancing binding is 
through the route of inducing a buckling of the graphene sheet.
We also consider C(VI) [see Fig.~\ref{fig2}(h)], which contains both -O- and -OH, but on the same side
of the sheet. While this is a higher energy configuration than that depicted in C(V), it leads to the highest binding strength of all the
configurations considered in this study, giving $E_{\rm ads}$ = 22.94 kJ/mol. This can be readily understood as a synergistic effect of the
hydroxyl and epoxy groups, since the electric fields of the hydroxyl and epoxy groups add together to result in a larger induced electric
dipole on the methane molecule.

\begin{figure}[]
\begin{center}
\includegraphics[width=6.0cm, height=5.8cm]{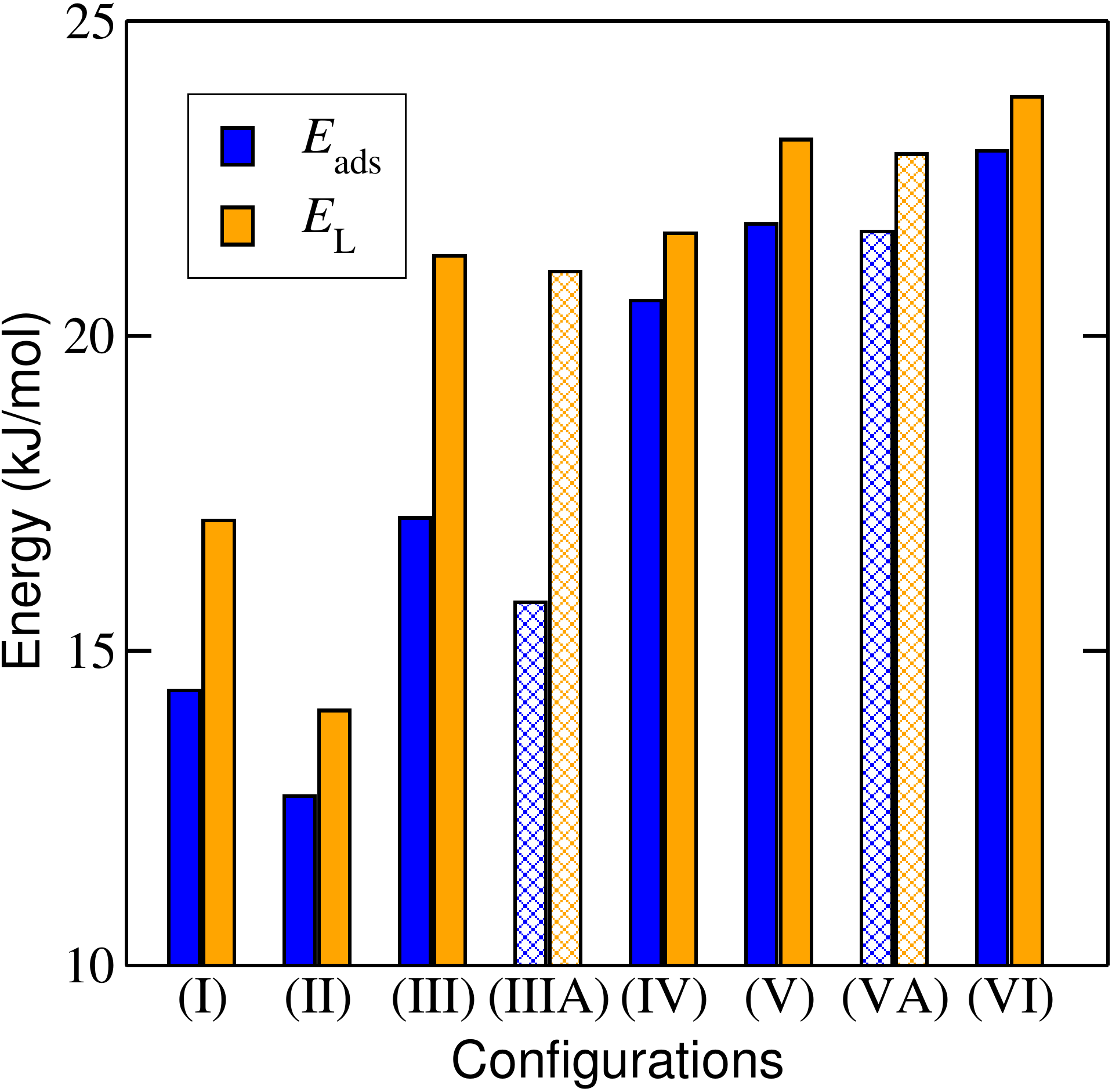}
\caption { Bar chart indicating step-wise enhancements in adsorption energy $E_{\rm ads}$, and the contribution $E_L$ to it from London dispersion interactions, for methane adsorbed on bare graphene (I) and a sequence of model systems (II)--(VI). See the text for a description of the systems. Configurations (IIIa) and (Va) are hypothetical in that they do not correspond to favorable configurations of the substrate, and are therefore shaded differently. Note that the solid blue bars increase at each stage, except for (II), which is energetically unfavored.}
\label{fig4}
\end{center}
\end{figure}

To summarize the results we obtain in this analysis of adsorption energies, we have plotted our results for $E_{\rm ads}$ and $E_L$, for the
various model configurations C(I) to C(VI), in the bar chart in Fig.~\ref{fig4}. Note that, unlike our results for the cases Z-OUTER, Z-INNER, and
A-OUTER, in all the cases considered here, $E_L$ is larger than $E_{\rm ads}$; this is because while $E_{\rm ads}$ contains attractive contributions from $E_L$, as well as from electrostatic interactions (which we will call $E_{ES}$), there are also repulsive contributions.
A small height difference between $E_{\rm ads}$ and $E_L$ means that $E_{ES}$ is large, as can be readily verified by comparing the heights
of the bars in Fig.~\ref{fig4} with the sizes of the red and green lobes in Figs.~\ref{fig1} and \ref{fig2}. With the exception of configurations C(IIIa) and C(Va), which were hypothetical, non-realistic configurations introduced for the purpose of separating out the effects of buckling from the electronic or charge-transfer effects due to -O-, we see that as we proceed from C(II) to C(VI), $E_{\rm ads}$ increases at every step. As we go from C(II) to C(III),
$E_{\rm ads}$ increases because $E_L$ increases; this in turn is because of the buckling of the graphene sheet, caused by the epoxy group. On proceeding from C(III) to C(IV), $E_L$ stays about the same; however, $E_{\rm ads}$ increases, because electrostatic attraction [which was practically
absent in C(III)] increases. C(V) differs from C(IV) in that there is an epoxy on the bottom side (opposite the methane); this results in
an additional buckling of the graphene sheet, as a result of which $E_L$ increases, and thus $E_{\rm ads}$ increases by a corresponding
amount. Finally, on going from C(V) to C(VI), $E_L$ increases slightly, but also the electrostatic interactions increase because of the
synergistic effect already discussed above, and thus we get the largest value of $E_{\rm ads}$. However, it is C(V) whose structure is both lower in energy as well as more representative of GO, and the value of $E_{\rm ads}$ in C(V) is similar to those obtained by us for the Z-INNER, Z-OUTER and A-OUTER geometries.

\section{Conclusion and Summary}
In conclusion, we have shown, by DFT calculations, that replacing graphene by graphene oxide increases the strength 
of binding to methane by about 50\%, which is sufficient to bring the adsorption energy into the target 
range for on-board vehicular storage. The enhancement comes from a synergy between various contributions to weak binding, 
viz., London dispersion forces and  electrostatic interactions, aided by structural distortion of the graphene sheet by epoxy groups. 
This suggests that graphene oxide and graphite oxide should be good candidate materials for the on-board storage of natural gas. 
The interfaces between patches of densely covered GO and graphene are
optimal binding sites for methane. Thus, if one could experimentally synthesize graphene oxide 
so as to deliberately engineer a high density of labyrinth-like interfaces between bare graphene and 
densely covered graphene oxide at the nanoscale, this should be optimal for
high-capacity methane storage.
Moreover, the understanding gained in this study can be leveraged to develop design 
principles that can be used to engineer future materials for methane storage, 
so as either increase the binding strength in nanocarbons even further 
beyond the values achieved here, or to transfer this knowledge to other materials.

\begin{acknowledgements}
Helpful discussions with Ganapathy Ayappa are gratefully acknowledged. Funding was provided by CSIR India, the DST Nanomission, and JNCASR.
\end{acknowledgements}


\begin{thebibliography}{99}


\bibitem{Dueren} T.~D\"uren, L.~Sarkisov, O.~M.~Yaghi and R.~Q.~Snurr, Langmuir \textbf{20}, 2683-2689 (2004).
\bibitem{Yaghi} M.~Eddaoudi, J.~Kim, N.~Rosi, D.~Vodak, J.~Wachter, M.~O'Keeffe and O.~M.~Yaghi, Science \textbf{295}, 469-472 (2002).
\bibitem{Mason} J.~Mason, M.~Veenstra and J.~R~Long, Chem. Sci. \textbf{5}, 32--51 (2014).
\bibitem{Liu} Y.~Liu, Z.~U.~Wang and H.-C.~Zhou, Greenhouse Gas Sci Technol. \textbf{2}, 239--259 (2014).
\bibitem{Lozano} D.~Lozano-Castell{\'o}, J.~Alca{\~n}iz-Monge, M.~A.~de la Casa-Lillo, D.~Cazorla-Amor{\'o}s and A.~Linares-Solano, Fuel \textbf{81}, 1777--1803 (2002).
\bibitem{Vidali} G.~Vidali, G.~Ihm, H.-Y.~Kim and M.~W.~Cole, Surface Sci. Rep. \textbf{12}, 135--181 (1991).
\bibitem{Bhatia-Langmuir} S.~K.~Bhatia and A.~L.~Myers, Langmuir \textbf{22}, 1688 (2006).
\bibitem{Yang} S.~Yang, L.~Ouyang, J.~M.~Phillips and W.~Y.~Ching, Phys. Rev. B. \textbf{73}, 165407 (2006).
\bibitem{Thierfelder} C.~Thierfelder, M.~Witte, S.~Blankenburg, E.~Rauls and W.~G.~Schmidt, Surf. Sci. \textbf{605}, 746--749 (2011).
\bibitem{Girifalco} L.~A.~Girifalco and M.~Hodak, Phys. Rev. B \textbf{65}, 125404 (2002).
\bibitem{Lu} D.~Lu, Y.~Li, D.~Rocca and G.~Galli, Phys. Rev. Lett. \textbf{102}, 206411 (2009).
\bibitem{Okamoto} Y.~Okamoto and Y.~Miyamoto, J. Phys. Chem. B  \textbf{105}, 3470--3474 (2001).
\bibitem{Zhao} J.~Zhao, A.~Buldum, J.~Han and J.~P.~Lu, Nanotechnology \textbf{13}, 195 (2002).
\bibitem{Wood} B.~C.~Wood, S.~Y.~Bhide, D.~Dutta, V.~S.~Kandagal, A.~Pathak, S.~N.~Punnathanam, K.~G.~Ayappa and S.~Narasimhan, J. Chem. Phys.  \textbf{137}, 054702 (2012).
\bibitem{Morris} R.~E.~Morris and P.~S.~Wheatley, Angew. Chem., Int. Ed. \textbf{47}, 4966--4981 (2008).
\bibitem{Wilcox} K.~Mosher, J.~He, Y.~Liu, E.~Rupp and J.~Wilcox, International Journal of Coal Geology \textbf{109}, 36--44 (2013).
\bibitem{Lua}  A.~C.~Lua, T.~Yang and J.~Guo, Journal of Analytical and Applied Pyrolysis \textbf{72}, 279--287 (2004).
\bibitem{Molina} M.~Molina-Sabio, C.~Almansa and F.~Rodr\'{i}guez-Reinoso, Carbon \textbf{41}, 2113--2119 (2003).
\bibitem{Kandagal} V.~Kandagal, A.~Pathak, K.~G.~Ayappa and S.~N.~Punnathanam, J. Phys. Chem. C \textbf{116}, 23394--23403 (2012).
\bibitem{Quinn} D.~F.~Quinn and J.~A.~MacDonald, Carbon \textbf{30}, 1097--1103 (1992).
\bibitem{Deb-JPCC} D.~Dutta, B.~C.~Wood, S.~Y.~Bhide, K.~G.~Ayappa and S.~Narasimhan, J. Phys. Chem. C \textbf{118}, 7741--7750 (2014).
\bibitem{Murali} Y.~Zhu, S.~Murali, W.~Cai, X.~Li, J.~W.~Suk, J.~R.~Potts and R.~S.~Ruoff, Advanced Materials \textbf{22}, 3906--3924 (2010).
\bibitem{Wang-2009} L.~Wang, K.~Lee, Y.-Y.~Sun, M.~Lucking, Z.~Chen, J.~J.~Zhao and S.~B.~Zhang, ACS Nano \textbf{3}, 2995--3000 (2009)
\bibitem{Lee-JNanoSci2013} S.-Y.~Lee and S.-J.~Park, J. Nanoscience and Nanotechnology \textbf{13}, 443--447 (2013).
\bibitem{Yildirim-JMC} G.~Srinivas, J.~W.~Burress, J.~Ford and T.~Yildirim, J. Materials Chemistry \textbf{21}, 11323--11329 (2011).
\bibitem{QE-JPCM2009} P.~Giannozzi, et al., J. Phys.: Condens. Matt. \textbf{21}, 395502 (2009). 
\bibitem{vanderbilt} D.~Vanderbilt, Phys. Rev. B \textbf{41}, 7892 (1990).
\bibitem{perdew1996} J.~P.~Perdew, K.~Burke and M.~Ernzerhof, Phys. Rev. Lett. \textbf{77}, 3865 (1996).
\bibitem{Grimme2007} S.~Grimme, J. Comp. Chem. \textbf{27}, 1787--1799 (2006).
\bibitem{marzari1999} N.~Marzari, D.~Vanderbilt, A.~De Vita and M.~C.~Payne, Phys. Rev. Lett. \textbf{82}, 3296 (1999).
\bibitem{Lerf1} A.~Lerf, H.~He, M.~Forster and K.~Klinowski, J. Phys. Chem. B (1998) \textbf{102}, 4477--4482 (1999).
\bibitem{Lerf2} H.~He, J.~Klinowski, M.~Forster and A.~Lerf, Chemical Physics Lett. \textbf{287}, 53--56 (1998).
\bibitem{Cai} W.~Cai, et al., Science \textbf{321}, 1815 (2008).
\bibitem{Erickson-AdvMater} K.~Erickson, R.~Erni, Z.~Lee, N.~Alem, W.~Gannett and A.~Zettl, Adv. Mater. \textbf{22}, 4467--4472 (2010).
\bibitem{Pacile-Carbon} D.~Pacil\'{e}, J.~C.~Meyer, A.~F.~ Rodr\'{i}guez, M.~Papagno, C.~G\'{o}mez-Navarro, R.~S.~Sundaram, M.~Burghard, K.~Kern, C.~Carbone and U.~Kaiser, Carbon \textbf{49}, 966--972 (2011).
\bibitem{Nguyen-PRB} M.-T.~Nguyen, R.~Erni and D.~Passerone, Phys. Rev. B \textbf{86}, 115406 (2012).

\end{thebibliography}
\end{document}